\documentclass[aps,twocolumn,notitlepage,prl,superscriptaddress,showpacs]{revtex4-1}

\usepackage{adjustbox}
\usepackage[utf8]{inputenc} 
\usepackage{hyperref}       
\usepackage{url}            
\usepackage{booktabs}       
\usepackage{amsfonts}       
\usepackage{nicefrac}       
\usepackage{microtype}      
\usepackage{subcaption}
\usepackage{bm}
\usepackage{braket}
\usepackage{graphicx}

\newtheorem{thm}{Theorem}

\usepackage{mathtools}
\usepackage{nccmath}
\usepackage{xcolor}
\usepackage{caption}
\usepackage{soul}

\newcommand{\Tr}{\text{Tr}}

\begin{document}
	
	\title{Efficient bipartite entanglement detection scheme with a quantum adversarial solver}
	
	\author{Xu-Fei Yin}
	\affiliation{Hefei National Laboratory for Physical Sciences at the Microscale and Department of Modern Physics, University of Science and Technology of China, Hefei 230026, China}
	\affiliation{Shanghai Branch, CAS Center for Excellence in Quantum Information and Quantum Physics, University of Science and Technology of China, Shanghai 201315, China}
	\affiliation{Shanghai Research Center for Quantum Sciences, Shanghai 201315, China}
	\author{Yuxuan Du}
	\affiliation{JD Explore Academy, Beijing 101111, China}
	\affiliation{School of Computer Science, Faculty of Engineering, University of Sydney, Sydney 2006, Australia}
	\author{Yue-Yang Fei}
	\affiliation{Hefei National Laboratory for Physical Sciences at the Microscale and Department of Modern Physics, University of Science and Technology of China, Hefei 230026, China}
	\affiliation{Shanghai Branch, CAS Center for Excellence in Quantum Information and Quantum Physics, University of Science and Technology of China, Shanghai 201315, China}
	\affiliation{Shanghai Research Center for Quantum Sciences, Shanghai 201315, China}
	\author{Rui Zhang}
	\author{Li-Zheng Liu}
	\author{Yingqiu Mao}
	\affiliation{Hefei National Laboratory for Physical Sciences at the Microscale and Department of Modern Physics, University of Science and Technology of China, Hefei 230026, China}
	\affiliation{Shanghai Branch, CAS Center for Excellence in Quantum Information and Quantum Physics, University of Science and Technology of China, Shanghai 201315, China}
	\affiliation{Shanghai Research Center for Quantum Sciences, Shanghai 201315, China}
	\author{Tongliang Liu}
	\affiliation{School of Computer Science, Faculty of Engineering, University of Sydney, Sydney 2006, Australia}
	\author{Min-Hsiu Hsieh}
	\affiliation{Hon Hai Quantum Computing Research Center, Taipei 114, Taiwan}
	\author{Li Li}
	\author{Nai-Le Liu}
	\affiliation{Hefei National Laboratory for Physical Sciences at the Microscale and Department of Modern Physics, University of Science and Technology of China, Hefei 230026, China}
	\affiliation{Shanghai Branch, CAS Center for Excellence in Quantum Information and Quantum Physics, University of Science and Technology of China, Shanghai 201315, China}
	\affiliation{Shanghai Research Center for Quantum Sciences, Shanghai 201315, China}
	\author{Dacheng Tao}
	\affiliation{JD Explore Academy, Beijing 101111, China}
	\affiliation{School of Computer Science, Faculty of Engineering, University of Sydney, Sydney 2006, Australia}
	\author{Yu-Ao Chen}
	\author{Jian-Wei Pan}
	\affiliation{Hefei National Laboratory for Physical Sciences at the Microscale and Department of Modern Physics, University of Science and Technology of China, Hefei 230026, China}
	\affiliation{Shanghai Branch, CAS Center for Excellence in Quantum Information and Quantum Physics, University of Science and Technology of China, Shanghai 201315, China}
	\affiliation{Shanghai Research Center for Quantum Sciences, Shanghai 201315, China}
	
	

\begin{abstract}
The recognition of entanglement states is a notoriously difficult problem when no prior information is available. Here, we propose an efficient quantum adversarial bipartite entanglement detection scheme to address this issue. Our proposal reformulates the bipartite entanglement detection as a two-player zero-sum game completed by parameterized quantum circuits, where a two-outcome measurement can be used to query a classical binary result about whether the input state is bipartite entangled or not. In principle, for an $N$-qubit quantum state, the runtime complexity of our proposal is $O(\text{poly}(N)T)$ with $T$ being the number of iterations. We experimentally implement our protocol on a linear optical network and exhibit its effectiveness to accomplish the bipartite entanglement detection for 5-qubit quantum pure states and 2-qubit quantum mixed states. Our work paves the way for using near-term quantum machines to tackle entanglement detection on multipartite entangled quantum systems.
\end{abstract}

\maketitle

\textit{Introduction.}---Quantum entanglement, the gem of quantum computation and  information processing, allows quantum computers to tackle certain problems beyond the traditionally possible. The representative examples include quantum phase estimation for factoring large integers~\cite{Shor:1997}, quantum cryptography for secure communication~\cite{Bennett:1992}, and quantum secret sharing~\cite{Hillery:1999}. In the last decade, tremendous progress has been achieved to fabricate various physical systems to manipulate entanglement~\cite{Song:2019, Monz:2011, Wang:2018, Zhong:2018}. Existing entanglement detection methods can be divided into three main groups. In the first group, the positive partial transposition method~\cite{Horodecki:1997} is generally efficient but only provides sufficient conditions for up to $2\times 3$ dimensional cases; the entanglement witness approach~\cite{Terhal:2000,Zhong:2018,Monz:2011} only distinguishes specific entangled states from separable ones; and the method based on self-testing~\cite{Harrow:2013} only effectively identifies product states. These methods can therefore only be regarded as “incomplete.” Conversely,  methods in the second group, such as symmetric  extension~\cite{Doherty:2004} and the  improved semidefinite programming hierarchy~\cite{Harrow:2017}, are complete in the sense that they can correctly identify any entangled state at any dimension, but the runtime cost is extremely expensive and exponentially scales with the qubits count. In addition to these deterministic  methods, classical machine learning techniques have also provided novel insights into the entanglement detection. For example, entanglement classifiers based on deep neural network~\cite{Roik:2021,Zhang:2018,Yang:2019,Gao:2018} can approximately recognize all entangled states with a high accuracy. Current learning algorithms oriented to entanglement tests belong to the supervised and static learning paradigm, where a labeled dataset containing thousands of quantum states is required to train classifiers. The adopted supervised learning techniques prohibit their scalability, since labeling and processing quantum states involving a large number of qubits is impractical on classical devices. Moreover, it has been recently shown that static algorithms are susceptible to adversarial attack and consequently false classifications, limiting their applicability~\cite{Lu:2020}. The above observations imply that there is currently no efficient solution to (approximately) identify the entanglement of  hundreds of qubit states due to the intrinsic computational hardness of the problem~\cite{Gharibian:2010,Gurvits:2003} and the exponentially large Hilbert space. Hence,  the certification of high-dimensional entanglement is a major challenge in quantum information processing~\cite{Friis:2019}.

Here we devise an efficient scheme to tackle  entanglement detection with scalability. Our proposal, the quantum adversarial bipartite entanglement detection scheme, is versatile and only requires modest resources, such that it can be efficiently built on noisy intermediate-scale quantum (NISQ) machines~\cite{Preskill:2018}. Furthermore, the proposed scheme is scalable and can be used to approximately recognize all quantum entangled states represented by any arbitrarily large qubit count, since the required computational complexity polynomially scales with the number of qubits. Mathematically, given an $N$-qubit state, the runtime complexity of our proposal is $O(\text{poly}(N)T)$, where $T$ refers to the total number of iterations. To our best knowledge, our theoretical study for the first time accomplishes high-dimensional entanglement testing of thousands of qubits with an acceptable computational overhead. Furthermore, our proposal benefits from robustness to adversarial attack, ensuring its practical applicability and reliability.  Our framework represents a new approach that uses NISQ devices to tackle practical quantum information processing problems with exponential advantages.    

\begin{figure*}[ht!]
	\centering
	\includegraphics[width=0.85\textwidth]{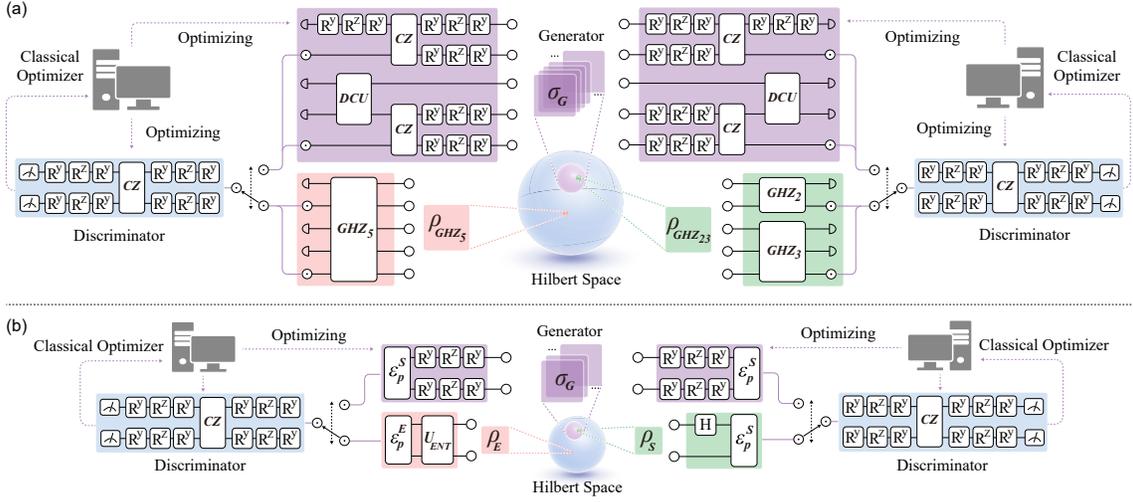}
	\caption{\small{\textbf{The implementation of the quantum adversarial bipartite entanglement detection scheme based on quantum circuit models}. (a) The bipartite entanglement detection between pure state $\rho_{G_{5}}$ and $\rho_{G_{23}}$ is shown in the left and right panels, respectively. In particular, the pink and green regions separately represent the state preparation of $\rho_{G_5}$ and $\rho_{G_{23}}$~\cite{Gate}. The purple region represents generator $U_G$, which is used to prepare $\sigma_{G}$ or $\sigma_{G}$, the blue region represents discriminator $U_D$ which is used to create a hyperplane that lies in the pure state space. (b) bipartite entanglement detection between mixed state $\rho_E$ and $\rho_S$ is shown in the left and right panels, respectively. Analogous to (a), the pink and green regions represent the state preparation protocols of $\rho_E$ and $\rho_S$. The symbols  $\it{H}$ ($U_{\text{ENT}}$) represents Hadamard(Entangled) gate used to prepare the pure state $\ket{\psi_{S}}$ ($\ket{\psi_{E}}$). $\mathcal{E}_p^{(S)}$ ($\mathcal{E}_p^{(E)}$)  represents the quantum noisy channel. The purple and blue regions represent the generator $U_G$ and discriminator $U_D$, which is used to prepare $\sigma_G$ and create a hyperplane that lies in the mixed state space. For both (a) and (b), at each iteration, the generated state and the given state are separately interacting with $U_D$ followed by a two-outcome POVM. The classical optimizer exploits the measured results to update trainable parameters in $U_G$ and $U_D$. }}
	\label{fig1}
\end{figure*}

\textit{The quantum adversarial bipartite entanglement detection scheme.}---Before describing the implementation details, we first introduce how to reformulate the bipartite entanglement detection as a two-player zero-sum game~\cite{Kale:2007}. This reformulation represents a precondition for devising efficient  quantum algorithms for bipartite entanglement detections. Recall that all quantum states form a convex set and all separable states form a convex subset~\cite{Nielsen:2010}. The convex property implies that given any separable-entangled state pair, there always exists a hyperplane that partitions these two states on opposite sides. This geometric observation allows us to set up a two-player zero-sum game~\cite{Goodfellow:2016,Farina:2017,Lloyd:2018}. Specifically, given an unknown state $\rho$,  the first player, i.e., the generator $G$, can only generate a separable state $\sigma_G$ located in the convex subset, while the second player, i.e., the discriminator $D$, can arbitrarily  create a hyperplane that lies in the quantum state space. The aim of $D$ is to discriminate $\sigma_G$ from $\rho$ by finding a hyperplane that places $\sigma_G$ and $\rho$ on opposite sides, while the aim of $G$ is to produce $\sigma_G$, which approximates  $\rho$ to fool $D$ in the sense that $\sigma_G$ and $\rho$ stay on the same side.  Remarkably, the restricted power of $G$ means that the Nash equilibrium, the status in which no player can improve their individual gain by choosing a different strategy, can be conditionally achieved if and only if $\rho$ is separable~\cite{Kale:2007}. This result converts the bipartite entanglement detection into a binary outcome optimization task, i.e., the achievable or unachievable equilibrium  correspond to  $\rho$ being separable or entangled.

We next elaborate on the quantum adversarial bipartite entanglement detection scheme, which accomplishes the zero-sum game under the NISQ setting. Specifically, denoted $\rho\in\mathbb{C}^{2^N\times 2^N}$ as a bipartite state that is composed of two subsystems $A$ and $B$. The task is to identify whether there exists entanglement between these two subsystems of $\rho$. To achieve this goal, the  generator $G$ and the discriminator $D$ used in our proposal refer to two trainable unitaries $U_G(\bm{\theta})$ and $U_D(\bm{\gamma})$, which are implemented by parameterized quantum circuits~\cite{Benedetti:2019,Du:2018}. Mathematically, we have $U_G(\bm{\theta})=\prod_{l=1}^{L_1}U_l(\bm{\theta})$ and  $U_D(\bm{\gamma})=\prod_{l=1}^{L_2}U_l(\bm{\gamma})$, where $L_1$ ($L_2$) refers to the number of blocks in $U_G(\bm{\theta})$  [$U_D(\bm{\gamma})$] and each block $U_l(\bm{\theta})$ [$U_l(\bm{\gamma})$] has an identical arrangement of quantum gates. The generated state is $\sigma_G= \Tr_E(U_G(\bm{\theta})(\ket{0}\bra{0})^{\otimes N'}U_G(\bm{\theta})^{\dagger})$, where $N'=N+N_A$, the subscript ``A'' represents the ancillary system that is formed by $N_A$ qubits, and $\sigma_G$ is obtained by partial tracing this ancillary system. When $\rho$ is a pure state, it is sufficient to set $N_A=0$. Note that to ensure that $U_G(\bm{\theta})$ can only prepare the states that are always separable between subsystems $A$ and $B$,  we impose a restriction on $U_l(\bm{\theta})$, where no two-qubit gate is allowed to span subsystems $A$ and $B$. The discriminator $D$ refers to a trainable two-outcome  positive operator valued measure $M_D$, i.e., $M_D  = U_D(\bm{\gamma})
	(\mathbb{I}\otimes E)U_D(\bm{\gamma})^{\dagger}$, where $\mathbb{I}\otimes E$ is the measurement operator with $E=\ket{0}\bra{0}$. In the above notation, the loss function  of the zero-sum game yields
\begin{equation}\label{eqn:5} 
	\max_{ \bm{\theta} }\min_{\bm{\gamma}} \mathcal{L}(\rho,\sigma_G)= \frac{
		\Tr(M_D\rho)}{2} + \frac{\Tr(M_D^{\perp}\sigma_G)}{2},
\end{equation}
where $M_D^{\perp}=\mathbb{I}-M_D$. The optimization of $\mathcal{L}$ amounts to iteratively updating $\bm{\theta}$ and $\bm{\gamma}$ with $T$ iterations. This updating process can be effectively completed by a classical optimizer (see Supplemental Material A and B in~\cite{SM}). The theoretical foundation of our proposal is guaranteed by the following theorem, whose proof is provided in Supplemental Material C~\cite{SM}.

\begin{thm}\label{thm1}
The zero-sum formulated in Eq.~(\ref{eqn:5}) has the equilibrium value $1/2$ if and only if the given state $\rho$ is separable. In this case, the generated state is identical to the given state, i.e., $\sigma_G=\rho$.
\end{thm}

As highlighted by Theorem~\ref{thm1}, the criterion for identifying the entangled states on NISQ devices shows that: during $T$ iterations, if the loss $\mathcal{L}$ converges to $1/2$ within a tolerable error, the state $\rho$ will be labeled as separable; otherwise, the state $\rho$ will be labeled as entangled. The runtime complexity of our scheme is $O[(L_1+L_2)N']$. This computational efficiency has two main origins. First, our proposal does not encounter read-in and read-out bottlenecks. Moreover, the gradient information for each parameter and the evaluation of $\mathcal{L}$ can be achieved in $O(1)$ runtime, which indicates that the optimization of each iteration takes $O[(L_1+L_2)N']$ runtime (see  Supplemental Material A~\cite{SM}). Second, unlike prior supervised learning based entanglement classifiers, our proposal does not require a preprocessing procedure to prepare labeled quantum states that may take $O(2^N)$ runtime. The low runtime cost assures the scalability of our proposal. 
 
\textit{Experimental setup.}---To benchmark the performance of the proposed scheme, we use linear optical circuits and entangled photon pairs to demonstrate the bipartite entanglement detection for two $5$-qubit pure states and two $2$-qubit mixed states, respectively. 

The construction of these states is as follows.  The first pure quantum state to be recognized is partially separable $\rho_{G_{23}} = \ket{G_2}\bra{G_2} \otimes \ket{G_3}\bra{G_3}$, where $\ket{G_N} =\frac{1}{\sqrt{2}}(\ket{H}^{\otimes N}+\ket{V}^{\otimes N})$ denotes the $N$-qubits $\text{Greenberger-Horne-Zeilinger}$ state. The second pure quantum state to be recognized is fully entangled $\rho_{G_{5}} = \ket{G_5}\bra{ G_5}$. Note that following the convention of linear optics, the qubit state $\{0, 1\}$ is represented by the polarization degree of freedom $\{H, V\}$, where $H$ and $V$ refer to the  horizontal and vertical polarizations, respectively. Moreover, the first mixed quantum state to be recognized is separable $\rho_S = \mathcal{E}^{(S)}_p(\ket{\psi_{S}}\bra{\psi_{S}})$, where $\ket{\psi_{S}}=\frac{1}{\sqrt{2}}(\ket{HH}+\ket{VH})$ is the separable pure state and $\mathcal{E}^{(S)}_p: \ket{\psi_{S}}\bra{\psi_{S}} \rightarrow \ket{\psi_{S}}\bra{\psi_{S}} - (1-p)(\ket{H}\bra{V}+\ket{V}\bra{H})\otimes \ket{H}\bra{H}$ represents the noisy channel. The second mixed quantum state to be recognized is entangled $\rho_E = \mathcal{E}^{(E)}_p(\ket{\psi_{E}}\bra{\psi_{E}})$, where $\ket{\psi_{E}}=\frac{1}{\sqrt{2}}(\ket{HV}+\ket{VH})$ refers to the entangled pure state and $\mathcal{E}_p^{(E)}: \ket{\psi_{E}}\bra{\psi_{E}} \rightarrow p \ket{\psi_{E}}\bra{\psi_{E}} + (1-p)(\ket{HH}\bra{HH} + \ket{VV}\bra{VV})/2$ represents the noisy channel. 

Our protocol's circuit-based model to identify the bipartite entanglement of the pure (mixed) quantum states is shown in the upper (lower) panel of Fig.~\ref{fig1}. For the pure state case, the block number for  $U_G(\bm{\theta})$ and $U_{D}(\bm{\gamma})$ is set as $L_1=2$ and $L_2=2$. $U_G$ includes $15$ parameterized single-qubit gates, two controlled-Z (CZ) gates, and a $6$ parameterized deterministic controlled unitary gate. For the mixed state case, the  block number for  $U_G(\bm{\theta})$ and $U_{D}(\bm{\gamma})$ is set as $L_1=1$ and $L_2=2$. $U_G$ includes $6$ parameterized single-qubit gates. To ensure that the generator can only produce separable states, for the pure state case, there is no two-qubit gate between the first two and the last three qubits in $U_G$; for the mixed state case, $U_G$ only includes single-qubit gates without two-qubit gates. In addition, for both cases, the discriminator $U_D$ consists of $12$ parameterized single-qubit gates and a CZ gate. Furthermore, in the mixed state case, to guarantee the same noise environment between the generated state and the input state, the quantum channel $\mathcal{E}_p^{(S)}$ is also employed in the generator with $p=0.8$~\cite{State_preparation}.

\begin{figure*}[ht!]
	\centering
	\includegraphics[width=0.85\textwidth]{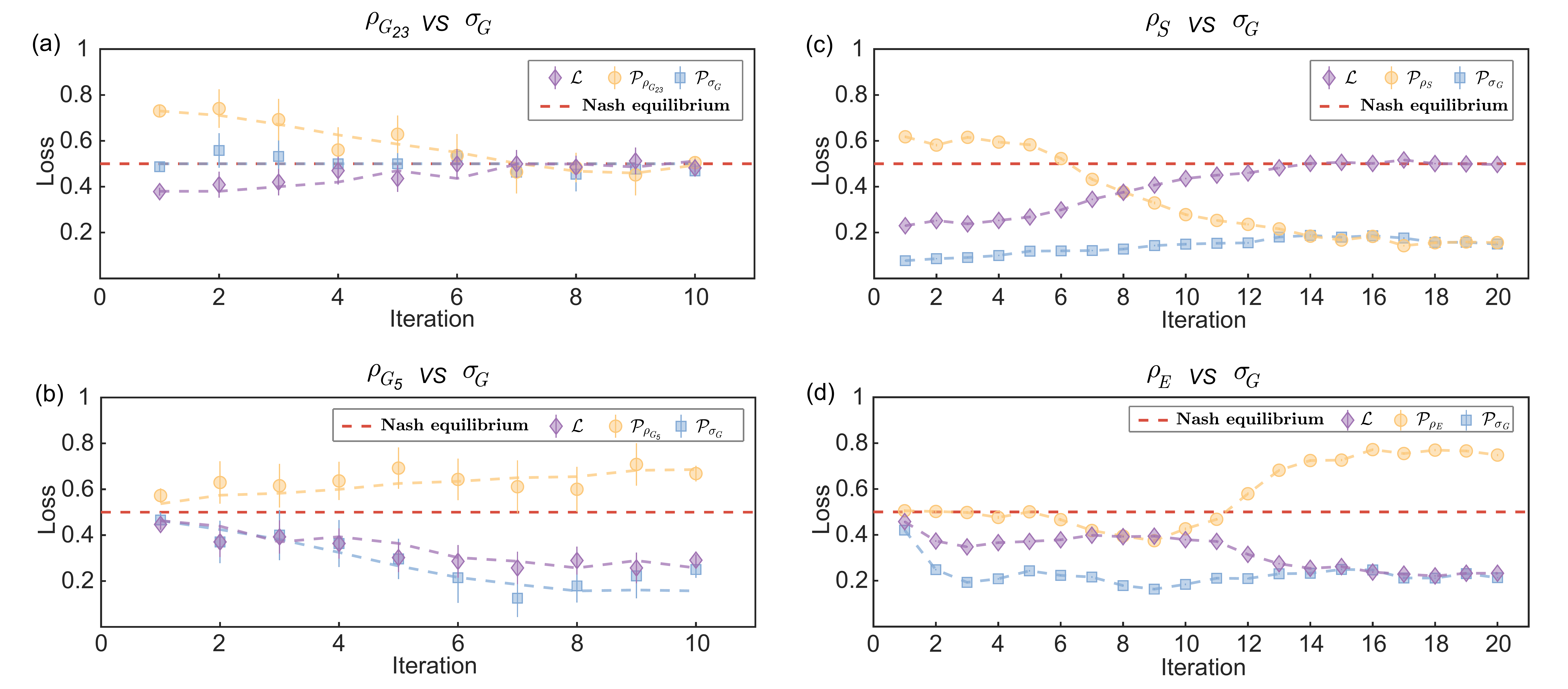}
	\caption{\small{\textbf{Experimental results for identifying the bipartite entanglement of $\rho_{G_{23}}$, $\rho_{G_{5}}$, $\rho_S$ and $\rho_E$.} The results between (a) $\sigma_{G}$ and $\rho_{G_{23}}$, (b) $\sigma_{G}$ and $\rho_{G_{5}}$, (c) $\sigma_{G}$ and $\rho_S$, (d) $\sigma_{G}$ and $\rho_E$. The labels `$\mathcal{L}$' and `$\mathcal{P}_{\sigma_{G}}$' refer to the loss $\mathcal{L}(\bm{\theta}^{(t)}, \bm{\gamma}^{(t)})$ and the terms $\Tr(M_D\sigma_{G}^{(t)})$ in Eq.~(\ref{eqn:5}). The labels `$\mathcal{P}_{\rho_{G_{23}}}$' , `$\mathcal{P}_{\rho_{G_{5}}}$', `$\mathcal{P}_{\rho_S}$' and `$\mathcal{P}_{\rho_E}$' refer to the terms $\Tr(M_D\rho_{G_{23}})$, $\Tr(M_D\rho_{G_{5}})$, $\Tr(M_D\rho_S)$ and $\Tr(M_D\rho_E)$ in Eq.~(\ref{eqn:5}), respectively. The red dash line indicates when loss $\mathcal{L}(\bm{\theta}^{(t)}, \bm{\gamma}^{(t)})$ converges to 0, `Nash equilibrium' is reached.
    }}
	\label{fig2}
\end{figure*}

\textit{Experiment results.}---The experimental identification of the entanglement of pure state $\rho_{G_{23}}$ is shown in Fig.~\ref{fig2}(a). The loss $\mathcal{L}(\bm{\theta}^{(1)}, \bm{\gamma}^{(1)})$ at the initial step is only $0.3784 \pm 0.0173$, which is far away from the Nash equilibrium of $0.5$. With an increased number of iterations, the loss quickly converges to the Nash equilibrium, i.e., $\mathcal{L}(\bm{\theta}^{(10)}, \bm{\gamma}^{(10)})$ reaches $0.4820 \pm 0.0180$ at the $10$th iteration. Moreover, the two terms $\Tr(M_D\sigma_{G}^{(10)})$ and $\Tr(M_D\rho_{G_{23}})$  tend to be equivalent, i.e., the state distance  $D(\sigma_{G}^{(10)},\rho_{G_{23}}) = \frac{1}{2}|\Tr(M_D\sigma_{G}^{(10)}) - \Tr(M_D\rho_{G_{23}})|$ equals to $0.0360 \pm  0.0284$. Given access to the reconstructed state, we obtain the fidelity $F(\rho_{G_{23}},\sigma_{G}^{(1)})$ is only $0.5299 \pm  0.0256$, while $F(\rho_{G_{23}},\sigma_{G}^{(10)})$ increases to $0.9202 \pm 0.0204$. These results echo with Theorem~\ref{thm1} for the separable case. 

The experimental results for identifying the entanglement of $\rho_{ G_5}$ are illustrated in Fig.~\ref{fig2}(b). In contrast to the partial separable case, the loss $\mathcal{L}$ for the state $\rho_{G_5}$, highlighted by the purple dashed line, converges to a point that is far away from the Nash equilibrium after $10$ iterations. Particularly,  although the loss $\mathcal{L}(\bm{\theta}^{(t)},\bm{\gamma}^{(t)})$ at the initial iteration is $0.4468 \pm 0.0230$, it decreases to $0.2908 \pm 0.0250$ at the $10$th iteration. Moreover, the distance $D(\sigma_{G}^{(10)},\rho_{G_{5}})$ is $0.1560 \pm  0.0280$.  We apply full tomographic measurements to reconstruct the states $\rho_{G_5}$ and $\sigma_{ G_5}^{(t)}$ with $t=1$ and $t=10$. The collected results show that the reconstructed state density matrix between $\rho_{ G_5}$ and $\sigma_{ G_5}^{(10)}$ is evidently disparate. By leveraging the reconstructed states, we obtain the fidelity at $t=1$ and $t=10$ are  $F(\rho_{ G_5},\sigma_{ G_5}^{(1)})= 0.2835 \pm 0.0214$ and $F(\rho_{ G_5},\sigma_{ G_5}^{(10)})= 0.3355 \pm 0.0333$, respectively. The above results accord with Theorem~\ref{thm1} for the entangled case.

We next apply our proposal to recognize the entanglement of the state $\rho_S$ and $\rho_E$.  The experimental identification of the entanglement of $\rho_S$ is shown in Fig.~\ref{fig2}(c). Specifically,  at the initial step, the loss $\mathcal{L}(\bm{\theta}^{(1)}, \bm{\gamma}^{(1)})$ is only $0.2299 \pm 0.0030$, which is far away from the Nash equilibrium of $0.5$. Analogous to the pure state case, the loss quickly converges to the Nash equilibrium, i.e., $\mathcal{L}(\bm{\theta}^{(20)}, \bm{\gamma}^{(20)})$ reaches $0.4969 \pm 0.0033$ at the $20$th iteration. Moreover, the two terms $\Tr(M_D\sigma_G^{(20)})$ and $\Tr(M_D\rho_S)$ tend to be equivalent, i.e., the experimental results lead to the distance  $D(\sigma_G^{(20)},\rho_S) =0.0031 \pm  0.0032$.  Through applying quantum state tomography~\cite{James:2001,Flammia:2012}, we further obtain the fidelity $F(\rho_S,\sigma_G^{(1)})$ is only $0.5546 \pm  0.0013$, while $F(\rho_S,\sigma_G^{(20)})$  increases to $0.9802 \pm 0.0004$.  The experimental results for identifying the entanglement of $\rho_E$ are illustrated in Fig.~\ref{fig2}(d). In contrast to the separable case, the loss $\mathcal{L}$ for the state $\rho_E$ converges to a point that is far away from the Nash equilibrium after $20$ iterations. In particular,  although the loss $\mathcal{L}(\bm{\theta}^{(t)},\bm{\gamma}^{(t)})$ at the initial iteration is $0.4578 \pm 0.0037$, it decreases to $0.2326 \pm 0.0028$ at the $20$th iteration with  $D(\sigma_G^{(20)},\rho_E) = 0.2675 \pm  0.0028$. By leveraging the reconstructed states, we obtain the fidelity at $t=1$ and $t=20$ iterations are  $F(\rho_S,\sigma_G^{(1)})= 0.4994 \pm 0.0011$ and $F(\rho_S,\sigma_G^{(20)})= 0.7094 \pm 0.0020$, respectively.  The above observations confirm the feasibility of our proposal in the mixed state scenario. Refer to Supplementary D~\cite{SM} for the omitted experimental results. 

\begin{figure}[ht!]
	\centering
	\includegraphics[width=0.43\textwidth]{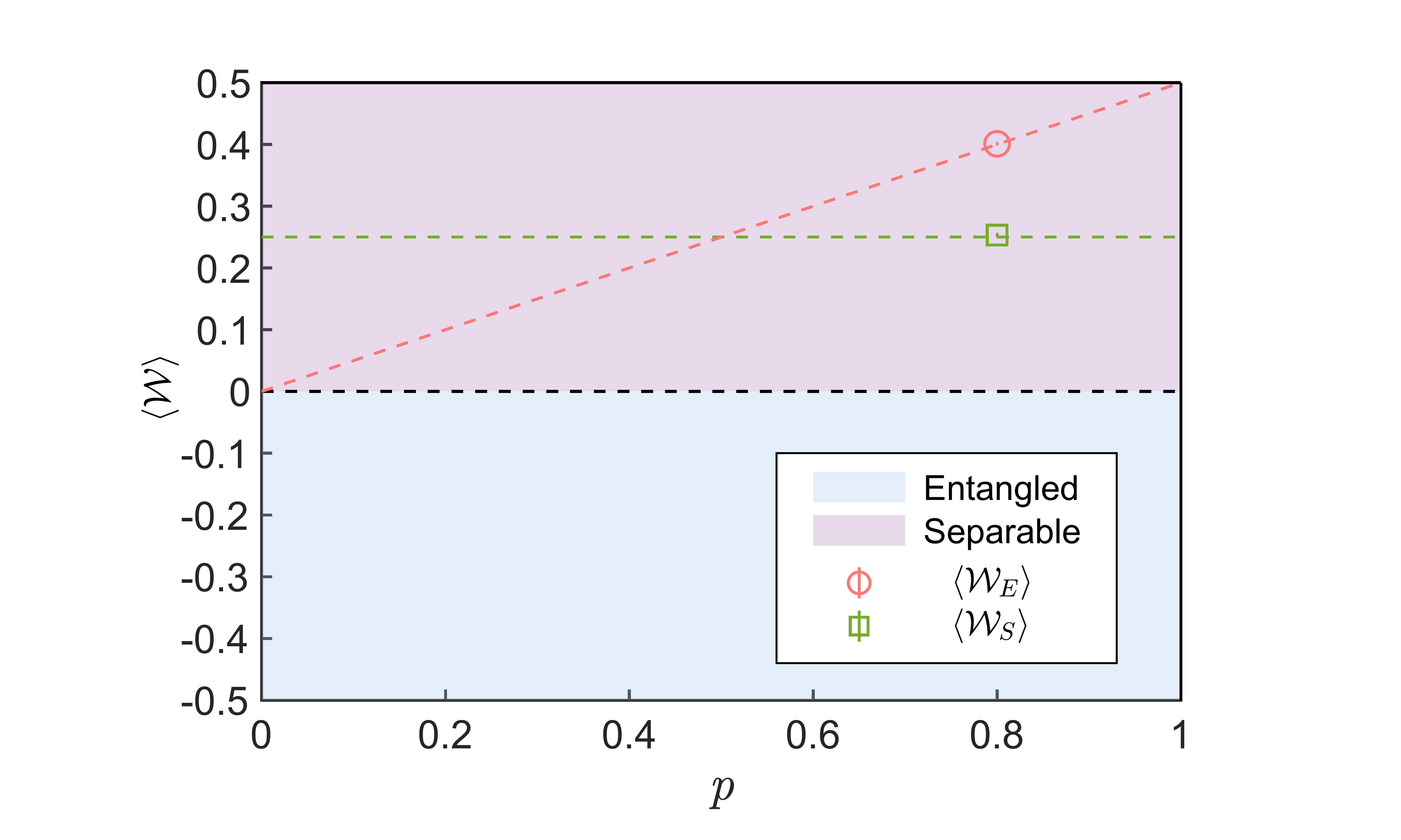}
	\caption{\small{\textbf{Entanglement witness result.}} The theoretical (experimental) results of using $\mathcal{W}_2$ to identify the entanglement of $\rho_S$ and $\rho_E$ with varying the depolarization rate $p$ ($p=0.8$) are labeled by green and red dashed lines (green square and red circle), respectively. }
	\label{fig3}	
\end{figure}

To exhibit that our proposal outperforms entanglement witness method~\cite{Terhal:2000} when no prior knowledge is available, we defer the comparison of our proposal with entanglement witness method to identify $\rho_S$ and $\rho_E$ in mixed scenario. Concretely, let $\mathcal{W}_N=\frac{1}{2}\mathbb{I}_{2^N}-\ket{G_N}\bra{G_N}$ be the employed multipartite entanglement witness operator, where $\mathbb{I}_{2^N}$ is the $2^N$-dimensional identity matrix. The criterion of entanglement witness methods is assigning the input state  $\kappa$ as entangled if $\Tr(\mathcal{W}_N\kappa)< 0$.  The experimental results of applying  $\mathcal{W}_N$ to identify $\rho_S$ and $\rho_E$  are shown in Fig.~\ref{fig3}. Specifically, the measured result of $\langle \mathcal{W}_E \rangle$ and $\langle \mathcal{W}_S \rangle$ for the state $\rho_E$ and $\rho_S$ is $0.4010 \pm 0.0007$ and $0.2532 \pm 0.0015$, which are both much larger than $0$. Consequently, the entanglement witness operator $\mathcal{W}_E$ provides a wrong prediction for $\rho_E$. 

\textit{Remark.} We heuristically estimate when the quantum adversarial bipartite entanglement detection scheme advances other complete methods and hence demonstrates potential quantum advantages. Also, our scheme can be effectively adapted to the fault-tolerant setting with provable convergence~\cite{Hazan:2016,Van:2019}. Concisely, for an $N$-qubit mixed state $\rho$, by setting the training rounds as $T$, our scheme completes the bipartite entanglement detection with the error $|\mathcal{L}(\bar{\sigma}_G,\rho) - 1/2|\leq 3\sqrt{N/T}$. Refer to Supplemental Material H and Theorem 2 in Supplemental Material I~\cite{SM} for details, respectively. Moreover, we conduct extensive experiments to analyze the robustness of our protocol. Given a set of 2-qubit mixed states, our protocol achieves $83.36 \%$ accuracy for identifying the bipartite entanglement in the measurement of the confusion matrix (See Supplemental Material E~\cite{SM} for details).

\textit{Conclusion.}---In this study, we have devised an efficient quantum adversarial bipartite entanglement detection scheme and experimentally demonstrated its efficacy on the linear optical network. Our proposal can be efficiently carried out on various quantum platforms. These properties are crucial in facilitating quantum technologies and understanding quantum mechanics.

In our proposal, the gate arrangements in $U_G$ and $U_D$ should be carefully designed, due to the tradeoff between the trainability and expressivity of the variational optimization algorithms~\cite{Jan:2019,Hu:2019,Huang:2020}. To further improve the performance of our scheme, a promising research direction is applying variable quantum circuit strategies~\cite{Grimsley:2019,Du:2020,Yao:2021} to automatically seek an optimal gates arrangement and maximize the benefit while minimizing the number of circuit elements. An alternative way is to exploit the realization of our proposal on fault-tolerant quantum chips, which possesses the proved convergence guarantee. Experimentally, we have demonstrated  the feasibility and efficiency of our bipartite entanglement detection scheme. Furthermore, we have compared our scheme with entanglement witness method in mixed case, and shown the reliability of our scheme when entanglement witness method failed. In future works, experimental comparison between our method and other complete entanglement detection methods is also worth exploring.  Moreover, we note that our scheme has potential to be extended to identify multipartite entanglement~\cite{horodecki2003entanglement,lu2018entanglement,Lu:2021,Chen:2021}. Although the objective function in Eq.~(\ref{eqn:5}) is generally hard to optimize, many recent studies proved that under certain assumptions, the gradient descent optimizers enable a good convergence rate~\cite{Yang:2020,Diakonikolas:2021,Ostrovskii:2021}. With this regard, an important future research direction is investigating whether these advanced techniques can accelerate the optimization of our proposal.

Finally, it is noteworthy that  a substantial class of tasks in the context of quantum information processing is explicitly quantifying entanglement of quantum states. The most exciting future work is extending our proposal to efficiently quantify unknown entanglement following these measures, which enables us to seek diverse potential applications of quantum computers with evident advantages. 

\bigskip
This work was supported by the National Natural Science Foundation of China (Grant No.~11975222), Shanghai Municipal Science and Technology Major Project (Grant No.~2019SHZDZX01), and Chinese Academy of Sciences and the Shanghai Science and Technology Development Funds (Grant No.~18JC1414700).

\bigskip
X.-F Y, Y.D., and Y.-Y. F. contributed to this work equally.

\end{document}